\documentclass[12pt]{article}
\usepackage[utf8]{inputenc}
\usepackage{amsmath,amssymb}
\usepackage{textcomp}
\usepackage{chet}
\usepackage{setspace}

\usepackage{hyperref}
\hypersetup{
	linktoc=all,
	colorlinks=false,
	hidelinks=true
}

\newcommand{\pr}{\prime}

\newcommand{\nn}{\nonumber}

\newcommand{\goesto}{\rightarrow}

\newcommand{\cL}{\mathcal{L}}

\newcommand{\cF}{\mathcal{F}}

\newcommand{\dif}{\mathrm{d}}

\newcommand{\link}[1]{[\href{http://arxiv.org/abs/#1}{{\tt arXiv:#1}}]}
\newcommand{\linkth}[1]{[\href{http://arxiv.org/abs/hep-th/#1}{{\tt arXiv/hep-th:#1}}]}
\newcommand{\linkgr}[1]{[\href{http://arxiv.org/abs/gr-qc/#1}{{\tt arXiv/gr-qc:#1}}]}
\newcommand{\doi}[1]{[\href{http://doi.org/#1}{{\tt doi:#1}}]}

\title{3D Lovelock Gravity and the Holographic c-theorem}

\author{Gökhan Alkaç$^{\ast,}$\email{alkac@mail.com} and
Gün Süer$^{\dagger,}$\email{suer.gun@metu.edu.tr}}

\affiliation{$^{\ast}$No Affiliation.\\
	$^{\dagger}$Department of Physics,\\
	Middle East Technical University, 06800 Ankara, Turkey.}

\abstract{We show that the scalar-tensor theory that arises in a rigorous $D \to 3$ limit of Lovelock gravity up to cubic order admits a holographic c-theorem and verify that the value of the c-function at the UV fixed point matches with the Weyl anomaly coefficient (up to an irrelevant factor). The constructed c-function is the same as one that follows from the ``naive limit'', which is obtained by scaling the couplings by a factor of $\frac{1}{D-3}$, and then setting $D=3$.}

\date{\today}

\begin{document}
\maketitle

\tableofcontents

\section{Introduction}

Given that gravity should have a geometrical interpretation due to the principles of equivalence and general covariance, it is possible to show the uniqueness of Einstein's theory in $D=4$ as follows \cite{Padmanabhan:2013xyr}: The structure of non-gravitational theories suggest that the metric tensor should be related to matter variables by a second-order differential equation. Therefore, the consistent coupling to a covariantly conserved energy-momentum tensor requires a symmetric, covariantly conserved tensor with at most second-derivatives of the metric tensor. Lovelock showed that the most general form of such a tensor in $D$-dimensions follows from the variation of an action functional 
\begin{equation}
	\label{L}
	S=\int \dif^D x\,\sqrt{-g} \,\mathcal{L}, \qquad \qquad \mathcal{L}=\sum_{m} c_m \mathcal{L}_m,
\end{equation}
where the $m$-th order Lovelock Lagrangian is given by \cite{Lovelock:1971yv,Lovelock:1972vz}
\begin{equation}
	\mathcal{L}_m=\frac{1}{2^m}\delta^{\mu_1\nu_1\cdots\mu_m\nu_m}_{\rho_1\sigma_1\cdots\rho_m\sigma_m}R^{\rho_1\sigma_1}_{\ \ \ \ \mu_1\nu_1}\cdots R^{\rho_m\sigma_m}_{\ \ \ \ \mu_m\nu_m}.\label{Lm}
\end{equation}
In addition to satisfying these basic requirements, Lovelock gravity has a unitary massless spin-2 graviton around any of its constant curvature vacua \cite{Sisman:2012rc} and is singled out among theories derived from a Lagrangian of the form $\mathcal{L}\,[g_{\mu\nu}, R_{\mu\nu\rho\sigma}]$ as the unique theory with equivalent metric and Palatini formulations \cite{Exirifard:2007da}. However; $\cL_{m}$ vanishes when $D<2m$ because of the anti-symmetrization in the generalized Kronecker delta symbol, and it is a topological invariant in $D=2m$,  i.e., the variation of the action yields a total derivative. Hence, a richer structure can only be achieved for $D>2m$, which leaves the Einstein-Hilbert action as the only possibility for having the above-mentioned properties in $D=4$.

Although this is somewhat disappointing for gravitational phenomenology, it is hard to overemphasize the importance of Lovelock gravity in testing theoretical ideas beyond Einstein's theory and the applications of the AdS/CFT correspondence \cite{Maldacena:1997re,Witten:1998qj,Gubser:1998bc}, our holographic framework to probe the properties of strongly coupled gauge theories, are no exception (see \cite{Brigante:2008gz,Brigante:2007nu,Buchel:2009sk,Camanho:2009vw,deBoer:2009pn,deBoer:2009gx,Camanho:2009hu} for examples). In this work, our focus will be the holographic c-theorem of \cite{Freedman:1999gp}, where it was shown for Einstein's theory in $D$-dimensions that it is possible to construct a monotonic function along an renormalization group (RG) flow induced by matter fields satisfying the null-energy condition (NEC). In the UV ($r \goesto \infty$), the value of the function is proportional to the Weyl anomaly coefficients of the boundary field theory, and as a result, one obtains a holographic generalization of Zamolodchikov's celebrated c-theorem for 2D quantum field theories (QFTs) ($c_\text{UV}>c_\text{IR}$ in any RG-flow connecting two fixed points) \cite{Zamolodchikov:1986gt} to arbitrary (even) dimensions. 

For a better understanding, a natural question needs to be answered: Although it is possible to construct a monotonic function along an RG-flow in generic $D$-dimensions, only even-dimensional conformal field theories (CFTs) have a Weyl anomaly. What should be the interpretation of the value of the function at the UV in the case of odd-dimensional CFTs? By studying Quasi-topological gravity (QTG) \cite{Myers:2010ru}, a particular higher-curvature gravity theory that allows distinguishing the type-A and type-B anomaly coefficients (associated with the Euler density and the independent Weyl invariants of weight $-d$ respectively), Myers and Sinha formulated a c-theorem in arbitrary dimensions	\cite{Myers:2010xs,Myers:2010tj}: Considering a $d$-dimensional CFT on $S^{d-1} \times R$, one can calculate the entanglement entropy of the ground state between
two halves of the sphere, and find a universal contribution. At the fixed points, the monotonic function is associated with this universal contribution ($a^*_\text{UV}>a^*_\text{IR}$) and for even $d$, as expected, it precisely matches with the A-type anomaly coefficient ($a^*=a$, and therefore $a_\text{UV}>a_\text{IR}$)\footnote{Note that the anomaly coefficient related to the Euler density is denoted by $c$ and $a$ for $d=2$ and $d\geq4$ respectively. Therefore, the monotonic function is denoted by $c(r)$ and $a(r)$, and the theorem is named as c-theorem and a-theorem accordingly. Throughout this paper, we keep using the former terminology even when we discuss the higher-dimensional cases since our main focus at the end will be 3D gravity theories with 2D CFT duals.}, providing further evidence for Cardy's conjecture in $d=4$ \cite{Cardy:1988cwa} that was later proven in \cite{Komargodski:2011vj}.

Based on the importance of higher-curvature interactions in establishing these results, in \cite{Sinha:2010ai}, Sinha investigated the analog of Lovelock Lagrangians \eqref{Lm} that naturally admit a c-function but vanish in 3D except the Einstein-Hilbert term ($m=1$). He showed that demanding the existence of a simple holographic c-theorem up to quadratic curvature invariants leads to New Massive Gravity (NMG), a non-linear completion of the Fierz-Pauli theory describing a unitary
excitation of massive spin-2 gravitons around the constant curvature vacua \cite{Bergshoeff:2009hq,Bergshoeff:2009aq}. In \cite{Sinha:2010ai}, cubic and quartic invariants admitting a c-function were also found and a prescription for obtaining such curvature invariants of arbitrary order was given in \cite{Paulos:2010ke}. Despite having higher-derivative field equations unlike Lovelock gravity, NMG and its extensions exhibit many similar properties especially in holography, and later a limit relating these three- and higher-dimensional theories at the Lagrangian level was discovered  \cite{Alkac:2020zhg}. Due to these similarities and also their own interesting properties, there has been an ongoing investigation of the holographic properties of 3D higher-curvature gravities \cite{Gullu:2010st,Jatkar:2011ue,Gullu:2010vw,Alkac:2018whk,Bergshoeff:2021tbz,Bueno:2022lhf,Bueno:2022log}.

In this paper, we will take a different route motivated by the recent developments over the past two years that started with a proposal for obtaining lower-dimensional ($D\leq2m$) versions of Lovelock gravity by Glavan and Lin, which can be summarized as follows \cite{Glavan:2019inb}: inserting a $D$-dimensional metric ansatz into field equations of Lovelock gravity gives terms that are proportional to $D-p$ with $p\leq2m$ so that they vanish when $D =p$. However; by scaling the relevant couplings in \eqref{L} as $c_m \goesto \frac{c_m}{D-p}$ and then setting $D = p$, one can obtain non-trivial solutions. In their paper, they studied $D \goesto 4$ limit of Einstein-Gauss-Bonnet theory ($m=1,2$ and $c_2$ is scaled by $\frac{1}{D-4}$) and obtained various novel solutions (constant curvature spacetimes, the
cosmological spacetimes, the spherically symmetric black holes, and the linearized fluctuations
around maximally symmetric vacua). Although this ``naive limit'' was later shown to be inconsistent on different grounds \cite{Gurses:2020ofy,Gurses:2020rxb,Arrechea:2020evj,Arrechea:2020gjw,Ai:2020peo,Fernandes:2020nbq,Mahapatra:2020rds,Bonifacio:2020vbk,Aoki:2020lig,Shu:2020cjw,Hohmann:2020cor,Cao:2021nng}, it led to the realization that rigorous procedures giving rise to well-defined theories exist \cite{Fernandes:2020nbq,Lu:2020iav,Kobayashi:2020wqy,Hennigar:2020lsl,Alkac:2022fuc} and they result in scalar-tensor theories with two-derivative field equations, i.e., examples of of Horndeski gravity \cite{Horndeski:1974wa,Kobayashi:2019hrl} or generalized Galileons \cite{Deffayet:2009mn,Charmousis:2012dw}. Properties of this special class of scalar-tensor theories have been investigated in \cite{Ma:2020ufk,Hennigar:2020drx,Hennigar:2020fkv,Khodabakhshi:2022knu,Mao:2022zrf}. 

Here we aim to find out whether the 3D scalar-tensor theory arising in a rigorous limit of Lovelock gravity up to cubic order, which we will refer to as 3D Lovelock gravity, admits a holographic c-theorem, and therefore constitutes an alternative to NMG and its extensions in this context. Although, as we show in Section \ref{sec:naive}, that the naive limit suggests the existence of a monotonic c-function along an RG-flow, it is known that some solutions do not survive in the well-defined scalar-tensor theories. For example, it was shown in \cite{Alkac:2022fuc} that 4D static black holes with spherical and hyperbolic horizons are excluded in 4D cubic Lovelock gravity. Therefore, this is a nontrivial check that needs to be performed for the evaluation of this theory in the holographic context.

The outline of this paper is as follows: In Section \ref{sec:naive},  we construct the holographic c-function and find its value at the UV fixed point of Lovelock gravity in $D>6$. Then, we present the results obtained from the naive limit. Section \ref{sec:3D} is devoted to the study of holographic c-theorem in 3D Lovelock gravity up to cubic term. We show that the theory admits a holographic c-function, and this function and its value at the UV fixed point match with the ones obtained from the naive limit. We also make a comparison with an earlier study of Horndeski gravity involving up to only linear curvature terms \cite{Li:2018kqp}. In Section \ref{sec:comment}, we comment on some important relations discovered in the study of 3D higher-derivative theories and their applicability in 3D Lovelock gravity. We end our paper with a summary of our results in Section \ref{sec:sum}.

\section{c-theorem in Lovelock Gravity in $D>6$ and Its Naive Limit to $D=3$}\label{sec:naive}
We will construct a holographic c-function for the cubic Lovelock gravity in $D>6$ by the method of \cite{Li:2017txk}. The action that we consider  is as follows:
\begin{equation}
	S=\int\dif^Dx\sqrt{-g}\left[R -2\Lambda_0+\alpha L^2\mathcal{L}_{m=2}+\beta L^4\mathcal{L}_{m=3}-\frac{1}{2}\partial_\mu\chi\partial^\mu\chi\right],\label{act}
\end{equation}
where the Lagrangians $\mathcal{L}_{m=2,3}$ can be found from the general expression \eqref{Lm}, and the free-scalar $\chi$ is introduced to impose the NEC in a practical way. Considering the following domain-wall ansatz
\begin{equation}
	\dif s^2=e^{2A(r)} \eta_{a b}\dif x^{a}\dif x^{b} +e^{2B(r)}\dif r^2,\label{domain}
\end{equation}
where $\eta_{a b}$ is the Minkowski metric and the Latin indices run from $0$ to $(D - 2)$, and assuming a radial profile for the free-scalar, $\chi = \chi(r)$, one can show that the NEC reduces to the positivity of the radial kinetic energy of the scalar field, i.e., $\frac{1}{2}\chi^{\pr 2}\geq 0$. Using this configuration in the action \eqref{act} gives an effective action for the functions  $\left[A(r),B(r),\chi(r)\right]$. After finding the corresponding Euler-Lagrange equations, and then fixing the gauge by $B(r)=0$, an inequality of the following form can be obtained by eliminating the bare cosmological constant $\Lambda_0$ from the equations:
\begin{equation}
	\frac{1}{2 } \chi^{\prime 2}=\cF\left(A^{\prime}, A^{\prime \prime}, \ldots\right) \geq 0,\label{eq:NEQ}
\end{equation}
where $\cF$ is a polynomial of the derivatives of the function $A(r)$. If it does not contain more than second-order derivatives of the function $A(r)$, and  has the form
\begin{equation}
	\cF=\sum_{n}  a_n {A^{\prime}}^{2n}A^{\prime\prime}\geq0,\label{Fgeneral}
\end{equation}
a monotonically increasing c-function can be defined as \cite{Alkac:2018whk}
\begin{equation}
	c(r):=\frac{1}{({\ell_P{A^\prime})}^{(D-2)}} \sum_{n}  \frac{a_n}{2(n+1)-D} {A^{\prime}}^{2n}.
\end{equation}
While the monotonicity is evident for even $D$, it was also shown to be the case by construction for odd $D$  in \cite{Myers:2010tj}. For the value of the c-function at the UV fixed point, we set $A(r)=\frac{r}{L}$ and find 
\begin{equation}
	c=\left(\frac{L}{\ell_P}\right)^{(D-2)} \sum_{n=0}^{D-1} \frac{a_n}{2(n+1)-D}.  
\end{equation}
The application of this procedure to cubic Lovelock gravity in $D$-dimensions yields the following inequality
\begin{equation}
	\mathcal{F}=(D-2)\left[-1+2\alpha(D-3)(D-4)L^2A'^2-3\beta(D-3)(D-4)(D-5)(D-6)L^4A'^4\right]A^{\prime\prime}\geq 0,
\end{equation}
which leads to the monotonically increasing c-function 
\begin{equation}
	c(r)=\frac{1}{({\ell_P{A^\prime})}^{(D-2)}}\left[1-2\alpha(D-2)(D-3)L^2A'^2+3\beta(D-2)(D-3)(D-4)(D-5)L^4A'^4\right],
\end{equation}
whose value at the UV fixed point is given by
\begin{equation}
	c=\left(\frac{L}{{\ell_P}}\right)^{(D-2)}\left[1-2\alpha(D-2)(D-3)+3\beta(D-2)(D-3)(D-4)(D-5)\right].
\end{equation}
It is obvious that this whole structure allows a naive limit to $D=3$ by scaling the couplings as $(\alpha,\beta)\to\dfrac{1}{D-3}\ (\alpha,\beta)$ and then setting $D=3$. The resulting expressions are
\begin{align}
	\cF &=\left[-1-2\alpha L^2A'^2+18\beta L^4A'^4\right]A^{\prime\prime}\geq 0,\label{F}\\
	c(r)&=\frac{1}{{\ell_P{A^\prime}}}\left[1-2\alpha L^2A'^2+6\beta L^4A'^4\right],\label{cr}\\
	c&=\frac{L}{{\ell_P}}\left[1-2\alpha +6\beta\right].\label{c}
\end{align}
In the next section, we will show that these expressions are indeed preserved in 3D Lovelock gravity up to the cubic term. In 3D, the value of the c-function at the UV fixed point $c$ should be proportional to the Weyl anomaly coefficient and the central charge of the boundary CFT.   
\section{c-theorem in 3D Lovelock Gravity}\label{sec:3D}
3D cubic Lovelock gravity is described by the action
\begin{equation}
	S=\int\dif^3x\sqrt{-g}\left[R -2\Lambda_0+\alpha L^2\mathcal{L}_{m=2}+\beta L^4\mathcal{L}_{m=3}-\frac{1}{2}\partial_\mu\chi\partial^\mu\chi\right],\label{act3}
\end{equation}
where we again introduce the free-scalar $\chi$ to impose the NEC and $\cL_{m=2,3}$ terms arising after a rigorous limit read
\begin{align}
	\mathcal{L}_{m=2}
	&=	4G^{\mu\nu}\phi_\mu\phi_\nu-4X\Box\phi+2X^2\label{GB2}\\
	\mathcal{L}_{m=3}
	&=-48R^{\mu\nu}\phi_{\mu\nu}X-48R^{\mu\nu}\phi_{\mu}\phi_{\nu}X+24 RX\Box\phi+6RX^2\nn\\
	&+96\phi_{\mu\nu}\phi^{\mu}\phi^{\nu}\Box\phi+48\phi_{\mu\nu}\phi^{\mu\nu}\Box\phi-24\phi_{\mu\nu}\phi^{\mu\nu}X\nn\\
	&-144\phi_{\mu\nu}\phi^{\mu}\phi^{\nu}X-96\phi^{\mu}\phi^{\nu}\phi_{\mu\rho}\phi_{\ \nu}^{\rho}-32\phi^{\mu\nu}\phi_{\mu\rho}\phi_{\ \nu}^{\rho}\nn\\
	&-16(\Box\phi)^3+24X(\Box\phi)^2-24X^3,
\end{align}
with the following definitions: $\phi_\mu\equiv\partial_\mu\phi$, $\phi_{\mu\nu}\equiv\nabla_\mu\nabla_\nu\phi$ and $X\equiv\partial_\mu\phi\partial^\mu\phi$. The $\mathcal{L}_{m=2}$ term is found in \cite{Hennigar:2020fkv,Hennigar:2020lsl,Fernandes:2020nbq} with the ``Weyl trick" of \cite{Mann:1992ar} that was first used to obtain the $D \goesto 2$ limit of general relativity. A more general version was obtained in \cite{Lu:2020iav} by a regularized Kaluze-Klein reduction, where one gets additional terms breaking the shift symmetry of the scalar $\phi$ that are proportional to the curvature of the maximally symmetric internal space considered in the reduction. When the internal space is taken to be flat, these different approaches give the same result up to field redefinitions. The $\mathcal{L}_{m=3}$ term is found in \cite{Alkac:2022fuc} by using both methods. We prefer the version arising from a flat internal space for simplicity.

In order to establish a holographic c-theorem for this theory, we will use the method described in the previous section by also assuming a radial profile for the scalar field $\phi=\phi(r)$ in addition the free-scalar $\chi$ and employ the domain wall ansatz \eqref{domain} in 3D, which lead to an effective action for the functions $\left[A(r),B(r),\phi(r),\chi(r)\right]$. This time, one arrives at the following inequality
\begin{equation}
	\mathcal{F}(A',A'',\phi',\phi'')
	=-A''+\alpha\left(4A'^2\phi'^2+4\phi'^4-2\phi'^2(A''-2\phi'')-4A'\phi'(2\phi'^2+\phi'')\right)\geq0,\label{ineq1}
\end{equation}
which is not in the standard form \eqref{Fgeneral}. However; considering the Euler-Lagrange equation of the function $\phi(r)$, which is given by
\begin{align}
	0&=(A'-\phi')\Big[A' \left(2 \phi '^2 \left(\alpha +27 \beta  \phi ''\right)-\alpha  \phi ''-36 \beta  \phi '^4\right)+\phi ' \left(3 \phi '' \left(\alpha -30 \beta  \phi '^2\right)-2 A'' \left(\alpha -18 \beta  \phi '^2\right)\right)\nn\\
	&+A'^2 \left(36 \beta  \phi '^3-2 \alpha  \phi '\right)\Big],
\end{align}	
one sees that this equation has two branches, both of which are solved by $\phi'=A'$. Inserting this into the original inequality \eqref{ineq1}, we find exactly the inequality suggested by the naive limit \eqref{F}. Therefore,  the 3D cubic Lovelock gravity admits a holographic c-function. The c-function and its value at the UV boundary are as follows:
\begin{align}
	c(r)&=\frac{1}{{\ell_P{A^\prime}}}\left[1-2\alpha L^2A'^2+6\beta L^4A'^4\right],\\
	c&=\frac{L}{{\ell_P}}\left[1-2\alpha +6\beta\right],\label{c3}
\end{align}
which are the same expressions obtained from the naive limit (\ref{cr},\ref{c}).

The relation between the AdS$_3$ radius $L$ and the bare cosmological constant $\Lambda_0$ can be easily found by setting $\phi'=A'=\frac{1}{L}$ and $\chi=0$ in the Euler-Lagrange equations following from the effective action, which yields
\begin{equation}
	\Lambda_0=-\frac{1}{L^2}\left(1+\alpha-6\beta\right).\label{vac}
\end{equation}
With this at hand, we can show the agreement with the Weyl anomaly coefficient \cite{Henningson:1998gx} of the boundary CFT by considering the theory on $S^2$ with the following Euclidean metric \cite{Emparan:1999pm}
\begin{equation}
	\dif s^2=\frac{\dif r^2}{1+ \frac{r^2}{L^2}}+r^2 (\dif \theta^2+\sin^2\theta \dif \phi^2),
\end{equation}
which is a solution when $\phi(r)=\log\left(\frac{r}{L}\right)$. In the on-shell action, we are interested in the logarithmic divergence since other divergences are expected to be canceled by surface terms and appropriately chosen counter terms. The relevant part of the on-shell action is as follows:
\begin{equation}\label{weyl}
	S \propto \left[\Lambda_0+\frac{1}{L^2}(3-3\alpha+6\beta)\right]\ln \left(\frac{2R}{L}\right),
\end{equation}
where $R$ is the UV cut-off scale. Upon using the relation between the bare cosmological constant $\Lambda_0$ and the AdS$_3$ radius $L$ in \eqref{vac}, we see that the anomaly coefficient is proportional to the value of the c-function at the UV boundary \eqref{c3}, as expected.

Note that the c-theorem in Horndeski gravity involving up to only linear curvature terms was investigated in \cite{Li:2018kqp}. With our conventions, the action that the authors considered is as follows:
\begin{equation}
	\label{eqn1}
	S=\int \dif^D x\,\sqrt{-g} \,\mathcal{L}, \qquad \qquad \mathcal{L}=R-2\Lambda_0-\frac{1}{2}\left(\alpha X - \gamma\, G^{\mu \nu}  \phi_\mu \phi_\nu\right).
\end{equation}
For an AdS vacuum solution characterized by the effective cosmological constant $\Lambda_\text{eff}=-\frac{D(D-1)}{2L^2}$ where $L$ is the AdS radius, the effective kinetic term for the Horndeski scalar $\phi$ is
\begin{equation}
	\cL_\phi=-\frac{1}{2}(\alpha+ \gamma \Lambda_\text{eff})X.\label{Keff}
\end{equation}
They found that a monotonic c-function is admitted only at the critical point where the effective kinetic term \eqref{Keff} vanishes, i.e., the couplings are related by $\gamma=\frac{2 \alpha L^2}{D(D-1)}$. Therefore, having second-order field equations is indeed not enough for the existence of a c-theorem. A particular combination of Horndeski couplings are required, and we see that 3D Lovelock gravity is such an example.
\section{Comment on the Wald Formula and the Central Charge Formula of \cite{Saida:1999ec,Kraus:2005vz}}\label{sec:comment}

Having shown the existence of a holographic c-theorem, we will now briefly comment on two important relations that appeared in the study of 3D higher-derivative theories described by the Lagrangians of the form\footnote{Note that this is the most general possibility for a pure gravity theory since the Riemann tensor can be written in terms of the metric, the Ricci tensor and the Ricci scalar in 3D due to the vanishing of the Weyl tensor.} $\mathcal{L}\,[g_{\mu\nu}, R_{\mu\nu}]$. In \cite{Sinha:2010ai}, an interesting relation between the c-function and the Wald formula \cite{Wald:1993nt,Iyer:1994ys} was observed: Evaluation of  the integrand of the Wald formula for the metric ansatz \eqref{domain} in 3D with $B(r)=0$ yields
\begin{equation}\label{wald}
	\frac{1}{2}g_{tt}g_{rr}\frac{\partial\cL}{\partial R_{trtr}}=c(r)A'(r).
\end{equation}
Additionally, the central charge of the boundary CFT for higher-derivative gravity theories, which is proportional to the anomaly coefficient, can be computed by the formula \cite{Saida:1999ec,Kraus:2005vz}
\begin{equation}\label{riccicentral}
	c=\frac{L}{3\ell_P}\frac{\partial\mathcal{L}}{\partial R_{\mu\nu}}g_{\mu\nu},
\end{equation}
where this quantity should be evaluated for the AdS$_3$ solution. When these relations are applied to 3D Lovelock gravity described by the action \eqref{act3} (without the $\chi$-term that is introduced for imposing the NEC), they yield the following expressions
\begin{align}
	c(r)&=\frac{1}{{\ell_P{A^\prime}}}\left[1+6\beta L^4A'^4\right],\\
	c&=\frac{L}{{\ell_P}}\left[1+6\beta\right].
\end{align}
Although there is no reason to expect these relations to hold for an arbitrary scalar-tensor theory, we see that the contribution from the $\mathcal{L}_{m=3}$ term matches our previous results, while the $\mathcal{L}_{m=2}$ term gives no contribution, which is incorrect. Interestingly, the correct contribution arises from the following Lagrangian
\begin{align}
	\cL_{m=2}&=8R^{\mu\nu}\phi_\mu\phi_\nu+8R^{\mu\nu}\phi_{\mu\nu}+4\phi_{\mu\nu}\phi^{\mu\nu}+8\phi_{\mu\nu}\phi^\mu\phi^\nu\nn\\
	&-4R\Box\phi-2RX-4(\Box\phi)^2+2X^2,
\end{align}
which is indeed the immediate result that is obtained after the application of the Weyl trick or the regularized Kaluza-Klein reduction to the $\cL_{m=2}$ Lagrangian in $D$-dimensions. The Lagrangian given in \eqref{GB2} is obtained after integration by parts and the use of Bianchi identities. Recently, it was observed in \cite{Khodabakhshi:2022knu} that when the Lagrangian holographic relation obeyed by Lovelock gravity in higher dimensions is demanded to be realized in the resulting scalar-tensor theories in lower dimensions, it requires introducing certain boundary terms by hand. Together with this finding, our results suggest that boundary terms play a crucial role in comparing Lovelock gravity in lower dimensions with pure gravity theories.

\section{Summary}\label{sec:sum}
In this paper, we studied the existence of a holographic c-function in 3D Lovelock gravity described by the Lagrangian \eqref{act3}, a scalar-tensor theory that arises after a rigorous $D\to3$ limit of Lovelock gravity. Being an example of a Horndeski theory, it has second-order field equations. However, as shown in \cite{Li:2018kqp} before, this does not guarantee the formulation of the holographic c-theorem. We have shown that the theory admits a holographic c-function and its value at the UV fixed point agrees with the one predicted by the Weyl anomaly in \eqref{weyl}. Moreover, our findings match the results obtained by the naive limit (\ref{F}-\ref{c}). By checking two important holographic relations \eqref{wald} and \eqref{riccicentral} of 3D higher-derivative gravity theories, we also shed light on the important role that boundary terms play in the comparison of lower-dimensional Lovelock gravity and pure gravity theories.	


\end{document}